# User Experience Design Professionals' Perceptions of Generative Artificial Intelligence

JIE LI, EPAM, Netherlands

HANCHENG CAO, Stanford University, United States

LAURA LIN, Google, United States

YOUYANG HOU, Notion Labs, United States

RUIHAO ZHU, Cornell University, United States

ABDALLAH EL ALI, Centrum Wiskunde & Informatica, Netherlands

Among creative professionals, Generative Artificial Intelligence (GenAI) has sparked excitement over its capabilities and fear over unanticipated consequences. How does GenAI impact User Experience Design (UXD) practice, and are fears warranted? We interviewed 20 UX Designers, with diverse experience and across companies (startups to large enterprises). We probed them to characterize their practices, and sample their attitudes, concerns, and expectations. We found that experienced designers are confident in their originality, creativity, and empathic skills, and find GenAI's role as assistive. They emphasized the unique human factors of "enjoyment" and "agency", where humans remain the arbiters of "AI alignment". However, skill degradation, job replacement, and creativity exhaustion can adversely impact junior designers. We discuss implications for human-GenAI collaboration, specifically copyright and ownership, human creativity and agency, and AI literacy and access. Through the lens of responsible and participatory AI, we contribute a deeper understanding of GenAI fears and opportunities for UXD.

## 1 INTRODUCTION

The landscape of Artificial Intelligence (AI) is rapidly evolving as adaptable foundation models, built on deep neural networks and self-supervised learning, have gained widespread adoption [6]. These span transformer-based large language models (e.g., GPT-3, BERT), visual (e.g., DALL-E, Florence), or multimodal (e.g., UniLM) models. The power of these models has recently been harnessed, exemplified by the launch of Generative AI (GenAI) tools, such as ChatGPT[1], DALL-E 2[2], Stable Diffusion[3], and Midjourney[4]. These models have dramatically pushed forward the edge of AI capabilities, where GenAI tools are capable of processing extensive data and learning statistical patterns, enabling them to generate novel multimodal output based on natural-language inputs (*i.e.*, prompts). They have demonstrated the ability to produce diverse types of content, including text [72], images [82], video [42], and audio [7]. For example, ChatGPT is viewed as the finest chatbot ever [36] with over a million people signed up to use it in only five days [79]. Enthusiastic fans posted examples of ChatGPT producing computer code, essays, and poems; while others type in surreal prompts to Midjourney or DALL-E 2, ranging from requests like"paint a portrait of your boss patting a tiger in the style of Rembrandt", to "depict a gathering of middle aged dinosaurs sipping coffee and contemplating the meaning of life", and these tools will return a startlingly accurate depiction in an instant. On the one hand, the widespread adoption of GenAI tools has sparked excitement within society due to their remarkable advancements; on the other hand, it results in anxiety about unanticipated consequences that may be implicitly induced and beyond human control [4]. Moreover, concerns arise regarding homogenization, as any weaknesses in these models have can propagate to downstream applications [6].

---

[1]ChatGPT: https://chat.openai.com
[2]DALL-E 2: https://openai.com/dall-e-2
[3]Stable Diffusion: https://stablediffusionweb.com
[4]Midjourney: https://www.midjourney.com

The expanding capabilities of GenAI have the potential to revolutionize various aspects of modern life, and significantly impacting individuals who earn their living by creating content, including designers, copywriters, journalists, and tenured professors. Among those professionals, there is a growing recognition of potential harms, such as reputation damage, economic and job loss, plagiarism, and copyright infringement [22, 25, 50, 75]. In the context of participatory AI, it is of utmost importance to recognize that creative professionals and content creators, not just technical experts, possess valuable knowledge, expertise, and interests crucial for the responsible development of AI [5]. Several efforts are underway to explore the impact of AI on UX (e.g., AI in UX Research report [10]). As a step to better understand and address such growing concerns, in this paper, we specifically explore the potential impact of GenAI tools on User Experience Design (UXD) practice, by involving UX Designers, important creative professionals, in our conversations. In contemporary UX Design, the goal is to align with both user needs and business values. Unlike other creative professions, UX Design is not purely artistic work, but revolves around meeting end users' demands. This entails not only designing the overall look and feel of a digital product [96], but also involves crafting effective navigation systems, labels, and content categorization to ensure that users can effortlessly locate and access the information they require [40]. The UX Design process is an iterative, user-centered design process that requires designers to be creative and collaborative, possess a problem-solving and sense-making mindset, and exhibit empathy towards users [73]. Successful UX Design must go through several iterations and is often linked to how well designers manage to comprehend and translate users' requirements into corresponding functionality and pleasing aesthetics [90]. With the ability to automate repetitive design tasks, such as creating wireframes, generating layout variations, or producing design prototypes [45], GenAI tools can significantly speed up the UX Design process, allowing UX Designers to explore a wider range of design options and iterate quickly to create more effective and engaging user experiences [81]. While GenAI tools offer many opportunities for UX Designers, it also presents potential threats related to job displacement, ethical concerns, lack of human touch, data quality and bias, and intellectual property issues [47].

## 2 RESEARCH QUESTIONS AND CONTRIBUTIONS

To harness the power of GenAI in supporting UX Designers and to ensure the responsible disclosure of further AI advancements, our objective is threefold. First, we aim to gain an overview of current UX Design practices, including the tools used, their limitations, as well as the existing workflows and challenges. Identifying existing practices and challenges provides a baseline against which the impact and utility of GenAI tools can be assessed. This also helps in contextualizing GenAI within current workflows. For instance, larger enterprises and smaller companies often operate differently in terms of UX design workflows. Understanding these contexts is crucial for evaluating how GenAI tools might fit into or potentially alter these various workflows. Second, we seek to comprehend UX Designers' perceptions toward GenAI – we intend to explore how UX Designers can leverage the potential of GenAI to overcome the existing limitations in tools and challenges in workflows, allowing them to ultimately craft impactful and meaningful user experiences now and in the future. Lastly, we seek to understand how and in what capacity GenAI may impact User Experience Design (UXD) practice, and whether fears that may arise are warranted. To this end, we pose the following research questions (RQs):

- **RQ1:** How do UX Designers perceive GenAI tools, and what is their view on the potential for incorporating these tools into their current workflows?
- **RQ2:** What opportunities and risks do UX Designers envision for the future of human-AI collaboration in UX Design practice?

To address these questions, we conducted one-on-one in-depth interviews with 20 UX Designers who have diverse years of experience and currently work in companies ranging from startups to large enterprises with over 10,000 employees. These companies are mainly located in four countries across Europe and the United States. We interviewed participants to characterize their practices, and inquire about their attitudes, concerns, and expectations.

Our work contributes to a deeper understanding of current UX Design practices, challenges as well GenAI fears and opportunities for UXD. We find that overall, experienced UX designers are confident in their skills in terms of originality, creativity, and user empathy skills. They believe that GenAI can serve as an assistive tool to help with repetitive and general tasks, and for enhancing productivity – however, only for those who are already skilled designers. They emphasized the unique human factors of "enjoyment" and "agency", where humans will remain as the final arbiters of "AI alignment" despite anticipations of future emergent AI abilities. However, they expressed serious concerns about skill degradation and unemployment, especially for junior designers, who may lose opportunities to systematically study design, and may instead end up being trained as prompters. Furthermore, they emphasized the importance of society paying sufficient attention to potential problems that may arise with such GenAI advancements, such as copyright and ownership issues, adverse impacts on human creativity ("creativity exhaustion") through GenAI output speed and homogenization, and concerns over promoting AI literacy and ensuring equal access to stay relevant. Ultimately, while UX Design practice may constantly be in flux, we believe GenAI's impact on this profession requires immediate adaptation. We discuss the implications of these findings, within the lens of ensuring responsible AI development and human-AI collaboration within UX Design practice.

## 3 BACKGROUND AND RELATED WORK

### 3.1 AI's Impact on (Creative) Work

A key line of related literature studies the impact of AI on work. Felton et al. [27] suggest that the top occupations exposed to language modeling include telemarketers and a variety of post-secondary teachers such as English language and literature, foreign language and literature, and history teachers. Eloundou et al. suggests [23] around 80% of the U.S. workforce could have at least 10% of their work tasks affected by the introduction of large language models, while approximately 19% of workers may see at least 50% of their tasks impacted. Solaiman et al. [89] describe specific social impact GenAI systems may have over society – the framework defines seven categories of social impact: bias, stereotypes, and representational harms; cultural values and sensitive content; disparate performance; privacy and data protection; financial costs; environmental costs; and data and content moderation labor costs. Other works have empirically shown the considerable productivity gains through the use of state of AI tools such as ChatGPT [71, 77].

A number of works have discussed the impacts of AI on creative works, which are closely relevant to UX design and research [32]. Pearson [76] delves into the role of modern creative AI technologies as potential muses for artists and creators, highlighting their capacity to enable novel forms of image creation, music composition, animation, and video production. Kulkarni et al. [55] investigates the use of text-to-image models (TTIs) in collaborative design through involving 14 non-professional designers. They reveal that TTIs facilitate rapid exploration of design spaces and support fluid collaboration, with text prompts acting as reflective design aids that facilitates exploration, iteration, and reflection in pair design. Ning et al.[69] investigate the design space of Artificial Intelligence technology-driven Creativity Support Tools (AI-CSTs), highlighting AI-CSTs' impact on workflows, their potential as co-creators, and strategies for handling AI errors, providing insights into AI-CSTs' design and technology requirements. Epstein et al. [25] examine the impact of generative AI tools on traditional artistic practices, addressing questions of authorship, ownership, and the potential transformation of creative work and employment. Chang et al. [15] studied the perception of artists who leveraged text-to-image AI models for artwork, finding that 1) artists hold the text prompt and the resulting image can be considered collectively as a form of artistic expression (prompts as art), and 2) prompt templates (prompts with "slots" for others to fill in with their own words) are developed to create generative art styles. Vinchon et al. [98] proposed four potential futures of AI human relationship for creative work, namely "Co-Cre-AI-tion", "Organic", "Plagiarism 3.0", and "Shut down". More closely related to the present work, Inie et al. [47] discussed creative professionals' worries and expectations about generative AI, concluding that creative professionals should better understand, cope with, adapt to as well as exploit AI. Long and Magerko [63] provided a concrete definition of AI literacy by identifying 16 core competencies (e.g., recognizing AI, human role in AI, ethics) that humans need to effectively interact with and critically evaluate AI, along with 15 recommended design considerations that foster increased user understanding of AI (e.g., explainability, promote transparency, low barrier to entry). Complementary to the foregoing, we focus and conduct in-depth analysis dedicated to understand how AI impacts one specific occupation closely related to the HCI community – UX Design practice.

### 3.2 Human-AI Collaboration for Human Empowerment

The relationship between human and machine intelligence has been discussed since the early days of human computer interaction [44, 60]. Given the increasing performance and prevalence of AI in society, much attention have been drawn to study and design for effective and responsible human-AI collaboration in recent years. Amershi et al. [2] proposed design guidelines for human-AI interaction. Through multiple rounds of evaluation with practitioners, they condense over 150 AI-related design recommendations into 18 aspects concerning four different time phases, and test their guideline against 20 AI-fused products with various categories (e.g., e-commerce, web search) and features (e.g., recommendations, search). Yang et al. [103] investigated whether, why, and how human-AI interaction exerts unique difficulties in design. Through synthesizing prior research with their own design, research, and teaching experiences, they identified AI capability uncertainty and AI output complexity as two sources of unique design challenges of AI. Over two decades ago, Horvitz [44] presented a set of principles for building mixed-initiative user interfaces (UIs) that enable users and intelligent agents to collaborate efficiently. The proposed principles are still widely applicable to today's human-AI collaboration, such as *"providing mechanisms for efficient agent-user collaboration to refine results"* and *"employing socially appropriate behaviors for agent-user interaction"*. Lehman [58] also investigates mixed-initiative human-AI interactions and collaborative work with generative systems. The author focuses on designing and evaluating functional prototypes through web-based experiments, exploring concepts like initiative, intent, and control, showing that the levels of initiative and control afforded by the UIs influence perceived authorship when writing text.

Kim et al. [52] used computational methods to categorize ten different AI roles prevalent in our everyday life and compared laypeople's perceptions of them using online survey data, e.g., AI are considered with roles such as tools (low in both human involvement and AI autonomy), servants (high human involvement and low AI autonomy), assistants (low human involvement and high AI autonomy), and mediators (high in both dimensions). They found that people assessed AI mediators the most favorably, and AI tools the least. Relatedly, Scott et al. [84] studied whether lay people perceive AI as "conscious", finding dynamic tensions between denial and speculation, thinking and feeling, interaction and experience, control and independence, and rigidity and spontaneity. Capel and Brereton [14] explored the emerging field of Human-Centered Artificial Intelligence (HCAI) by reviewing the existing literature on the subject. Their work identifies established HCAI clusters, and highlights emerging areas, including Interaction with AI and Ethical AI. Additionally, they proposes a new definition of HCAI and encourage greater collaboration between AI and HCI research while suggesting new HCAI constructs. Similarly, Shneiderman [86] explored the concept of HCAI as a transformative approach to AI system design, emphasizing the importance of placing humans at the core of design thinking, offering three key ideas: (1) a two-dimensional HCAI framework that balances human control and automation, (2) a shift from emulating humans to empowering people, and (3) a three-level governance structure for creating trustworthy HCAI systems. These ideas propose a reframe in design discussions with the potential to yield greater benefits for individuals, communities, and society, although they will require validation and refinement through further research and development. Indeed, such HCAI methods [85] aim to ultimately produce Reliable, Safe & Trustworthy (RST) designs, fostering human self-efficacy, creativity, and responsibility while significantly improving performance. Birhane et al. [5] reviewed participatory methods and practices within the AI and Machine Learning pipeline and acknowledged that wider publics beyond technical experts have knowledge, expertise, and interests that are essential to the design, development, and deployment of AI. These prior works set the stage for ensuring, at both a research and policy level, for ensuring responsible human-AI collaboration that ultimately empower people and practices. We draw on these works to help interpret our findings on how UX Designers envision current and future human-AI collaboration.

### 3.3 Use of AI in UX Design and Research

Studying User Interface (UI) and corresponding User Experience (UX), and how to build tools to better support UX design and research has long been a focal subject in the HCI community [12, 32, 53]. As AI increasingly shows its potential in transforming work practices, researchers have started studying and discussing the implications of AI for UX design and research [51, 90]. So far research has demonstrated the immense potential of AI for UX Design, with prototypes and applications available at key processes, including understanding the context of use, uncovering user requirements, aiding problem solving, evaluating design, and for assisting development of solutions [1, 90]. Generally, most believe that AI will likely result in AI-augmented creativity support tools rather than a full transformative shift [61, 98]. Tholander et al. [92] investigated how generative machine learning and large language models may play a part in creative design processes of ideation, early prototyping and sketching, demonstrating practical usefulness and limitations of the system in design ideation processes, as well as how user interaction, and broader discourse around AI, shapes user expectation of AI's capabilities and potentials. Feng et al. [28] found that UX practitioners can be more hands off when designing prototypes since AI can provide support. They introduced in detail how UX practitioners communicate with AI and use AI as design materials for their daily work. Other works have also proposed specific prototypes to support better user experiences [38, 99]. Evaluations of such systems point to the potential of AI for augmenting human creativity, as AI often brings distinct perspective that opens up new avenues for artistic expression[16], and fosters a two-way exchange of ideas between users and AI [56]. Lin et al. [62] explored the design space of mixed-initiative co-creativity systems where humans and AI systems could communicate creative intent to each other.

Meanwhile, research has also revealed challenges of incorporating AI into the UX design process. Gmeiner et al. [35] showed designers face many challenges in learning to effectively co-create with current generative AI systems, including challenges in understanding and adjusting AI outputs and in communicating their design goals. For example, when expecting AI tools to provide them with project-relevant work examples, or exhibit more context awareness for the specific problem at hand. Zhang et al. [106] further demonstrated limitations of AI for architectural design, in that the AI tool "fails to understand the Architectural domain-specific terms" and "generates surreal images unsuitable for construction purposes". Other works demonstrate similar challenges of prompt engineering in using AI for product design [43]. On the other hand, Abbas et al. [1] found that while the majority of UX designers had no expertise with ML as a design tool, it holds great potential for improving productivity. Thoring et al. [93] define a research agenda for GenAI-enabled design, highlighting key areas include creating better guided prompts, better interfaces for output interpolation, and getting AI to think outside of the latent space box (i.e., extrapolation). With regard to the potential of AI in UX research, Yang et al. [102] proposed a methodology to simulate user experience by using AI-aided design technology in mobile application design. Hamalainen et al. [37] leveraged OpenAI's GPT-3 model to generate open-ended questionnaire responses about user experiences over video games, where they find that GPT-3 can yield believable accounts of HCI experiences that is hard to be distinguished from real human responses. Park et al. [74] leveraged GPT-3 to simulate synthetic users and conversations over social computing platforms prototypes, where they find generated responses are hard to be differentiated from actual community behavior. By contrast to the majority of literature that focus on evaluating the use of AI for UX design and research in specific settings or technologies, our work contributes a deeper understanding of current practices, AI adoption, and perceptions of UX practitioners regarding emerging generative AI technologies, which sets the stage for envisioning how UX as a profession may evolve in the future with the ever increasing capabilities of AI.

## 4 METHOD

We employed a semi-structured interview methodology for this study [64]. We initially distributed a screening questionnaire with the goal of recruiting UX Designers who have diverse educational backgrounds and industry experiences, including years of experience, working at various company sizes, and residing in at least two different continents (namely Europe and North America). The screening questionnaire was distributed across social media networks (e.g., LinkedIn) and mailing lists (including design schools), targeting UXD communities. We received 48 responses, all of which were reviewed. While all respondents met our recruitment criteria, only 20 participants chose to participate. This mix of participants was well balanced in terms of our recruiting requirements. These 20 participants engaged with us in online one-on-one 60-minute interviews between April and July, 2023. 20 participants were deemed sufficient given saturation points exceeding 17 interviews in sociological research [41] and local sample size standards in HCI (remote interviews usually have M=15, SD=6) [13]. Each interview was conducted by an interviewer (author), while a second researcher (also author) was present to take notes. All interview sessions were additionally audio recorded. The interview process, data collection, and storage strictly adhered to the GDPR and ethical policy and data protection guidelines set forth by the last author's research institute.

| Characteristics | | UX Designers |
|---|---|---|
| **Work experience** | >5 years | P1, P5, P6, P8, P10, P11, P12, P14, P16, P17, P18 |
| | 3-5 years | P2, P7 |
| | <3 years | P3, P4, P9, P13, P15, P19, P20 |
| **Company size** | >10,000 | P5, P6, P7, P8, P9, P11, P14, P15, P16, P17 |
| | 1,000-10,000 | P1, P2, P10, P13, P18 |
| | 500-1,000 | P20 |
| | <50 | P3, P4, P12, P19 |

Table 1. Years of work experience and company size (employee count) for our 20 UX Designer participants (P1-P20)

| GenAI video demos | GenAI tools included | Date posted on YouTube |
|---|---|---|
| **Crafting a mobile interface**[5] | Midjourney (V4), Stability Photoshop Plugin (0.7.0) & ChatGPT (Jan. 30, 2023 version) | Feb. 20, 2023 |
| **Drafting a design proposal** (only used the part of Notion AI: 4'18"-4'50")[6] | Notion AI (2.21) | Mar. 15, 2023 |
| **Creating UI components**[7] | Prompt2Design (V3) | May. 6, 2023 |
| **Organizing ideation notes**[8] | MiroAI (Beta) | Mar. 8, 2023 |

Table 2. GenAI video demonstrations used in the interview

### 4.1 Participants

To ensure traceability in presenting the results, we labeled the 20 participants as P1-P20. Among the 20 participants, there were 12 females and 8 males. Three participants fell within the 18–25 age group, while 12 were in the 26–35 group. An additional five participants belonged to the 36–45 age group. 11 of them had over five years experience as UX Designers, two had three to five years, and seven had fewer than three years of experience. For their experience using GenAI tools, all 20 participants used *ChatGPT* regularly, and eight of them regularly used *GPT-4* (known as ChatGPT Plus). Thirteen participants tried *Midjourney* and occasionally used it for inspiration. Seven participants also mentioned their occasional use of *Notion AI*. Four tried *Miro AI*, and two used *Bard*.

The 20 participants were from 15 different companies, with ten participants located in Europe, including The Netherlands, Sweden, Romania, Hungary, and the other 10 participants located across the United States. Moreover, ten participants were employed by enterprises with over 10,000 employees (across well known Big Tech[9] companies), five worked in internationally known enterprises with 1,000 to 10,000 employees, one was employed at a company with 500 to 1,000 employees, and four worked in small companies or startups with fewer than 50 employees. To ensure data privacy, we ensure that where participants work cannot be traced back to them (Table 1).

### 4.2 Interview questions and procedure

The interview questions are scripted to gather answers for the research questions presented in Sec. 1, which are divided into four parts. **Part 1** begins with general questions that aim to gather information about participants' backgrounds, work experiences, common challenges in UX Design, and perceived limitations in their current tools. **Part 2** comprises a set of questions focused on understanding participants' current experiences and knowledge of GenAI tools, as well as their attitudes toward using GenAI in UX Design work. In **Part 3**, participants are shown a collection of videos consisting of four demonstrations from YouTube using various GenAI tools for UX Design, ideation, research data synthesis, and front-end development work (Table 2). We used these video demonstrations primarily as probes to gauge our participants to reflect on these tools in general and how they relate to their practices, and not to assess whether or not they have used these specific tools themselves. Subsequent follow-up questions delve into their impressions of the GenAI tools shown in the videos, their envisioned roles for AI in human-AI collaboration, and their concerns regarding ethics and ownership. **Part 4** concludes the interview with questions soliciting further reflections, advice or recommendations for UX Designers interested in integrating GenAI into their work and identifying their most desired, innovative functionalities of current and future GenAI. The complete list of interview questions is available in Supplementary Material A.

---
[9]https://en.wikipedia.org/wiki/Big_Tech

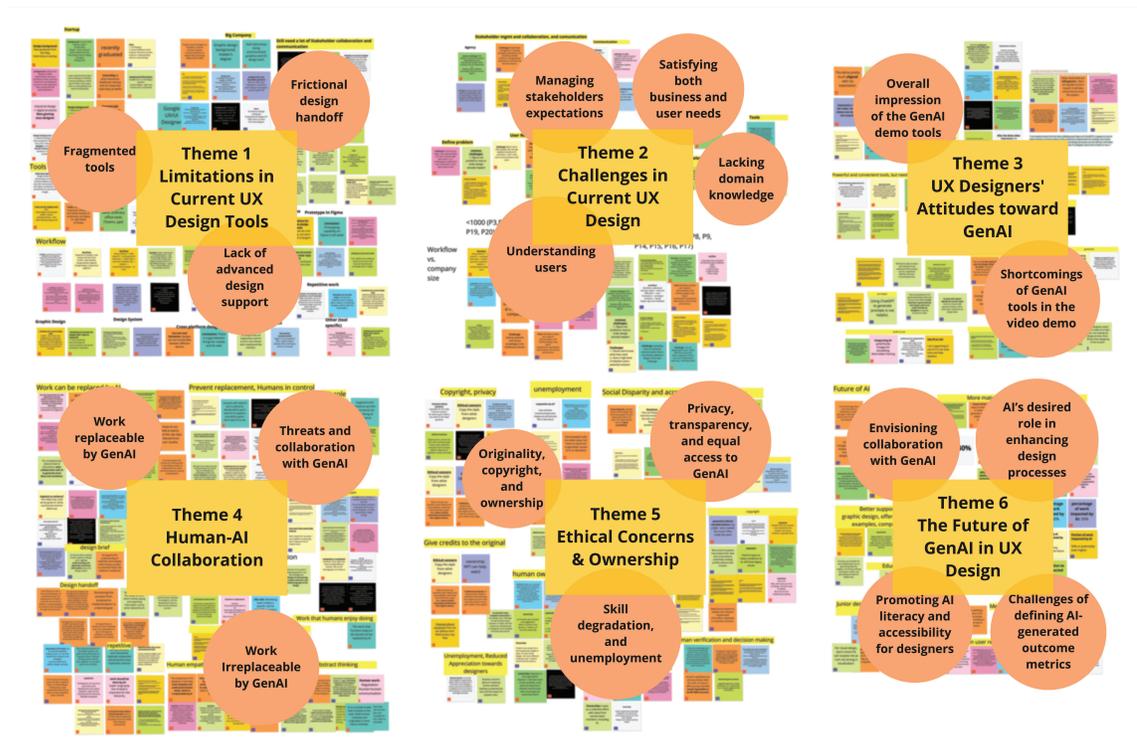

Fig. 1. The analysis resulted in six major themes, where each theme contained two to four sub-themes.

The interview process lasted approximately 60 minutes and involved the following steps:

- **Step 1 (5 mins):** Introduction given by the interview facilitator to ensure participants understand the purpose of the interview, have read, comprehended, and signed the informed consent form;
- **Step 2 (10 mins):** General questions about background, workflows, and UX practice;
- **Step 3 (20 mins):** Attitudes, concerns, and expectations of GenAI Systems;
- **Step 4 (20 mins):** Envisioning GenAI's role in human-AI collaboration (including watching a video demonstration of a sample of current GenAI tools);
- **Step 5 (5 mins):** Concluding questions.

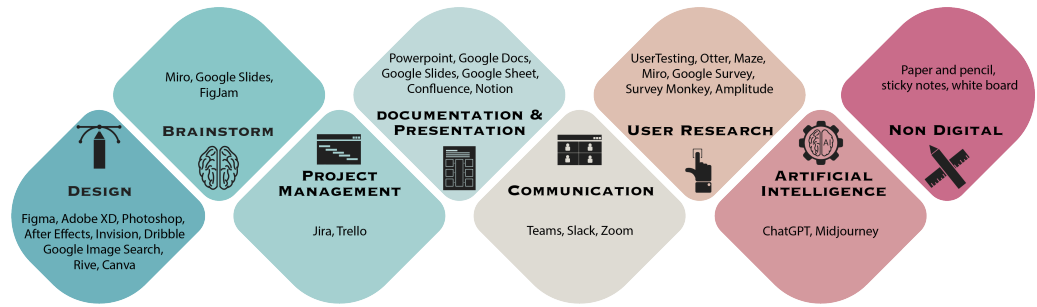

Fig. 2. Eight categories of tools commonly used in UX Design practices

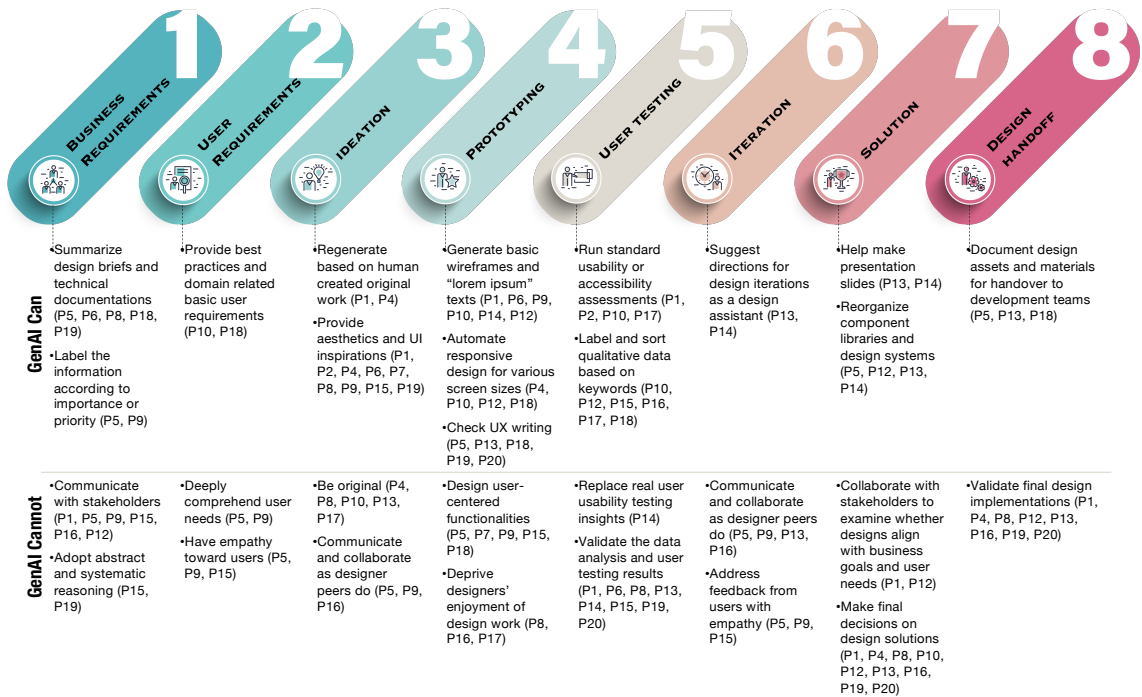

Fig. 3. Activities that can and cannot be completed by GenAI in the UX Design workflow

## 4.3 Data Analysis

We used Miro[10] for remote collaboration on data analysis. The audio recordings of the interviews were transcribed. We analyzed the data following a deductive/inductive hybrid thematic analysis approach [29, 101]: First, the transcripts were coded by five researchers using a deductive coding approach, where each researcher identified topics in the transcripts independently according to four categories in our code manual, created based on our research questions and interview guide. After the coding phase, the five researchers held a half-day workshop to compare and discuss the identified topics and converted the coded data into 415 digital statement cards on Miro. Each statement card consists of the original quote from the participant, participant ID (P1-P20), and a one-sentence statement or summary of the quote on the card. Instead of calculating statistical inter-rater reliability (IRR) for the analysis, the consensus among the five researchers was reached through daily meetings, workshops and discussions [67]. Given that the categories in the code manual were still too broad, we then adopted an inductive coding approach [9, 17] to surface sub-themes from the statement cards using the affinity diagramming technique [39], and to ensure the broader categories are consolidated. To help ensure our analysis meets the trustworthiness criteria (cf., [70]), we describe our hybrid approach in detail in Supplementary Material B. The analysis resulted in six major themes (Fig. 1): **(1) Limitations in current UX Design tools:** (a) frictional design handoff; (b) fragmented tools; and (c) lack of advanced design support. **(2) Challenges in current UX Design:** (a) managing stakeholders expectations; (b) understanding users; (c) satisfying both business and user needs; and (d) lacking domain knowledge. **(3) UX Designers' attitudes towards GenAI:** (a) overall impression of the videos; and (b) shortcomings of GenAI tools in the videos. **(4) Human-AI collaboration:** (a) threats and collaboration with GenAI; (b) work replaceable by GenAI; and (c) work irreplaceable by GenAI. **(5) Ethical concerns and ownership:** (a) originality, copyright, and ownership; (b) skill degradation, and unemployment; (c) privacy, transparency, and equal access to GenAI. **(6) The future of GenAI in UX Design:** (a) envisioning collaboration with GenAI; (b) promoting AI literacy and accessibility for designers; (c) AI's desired role in enhancing design processes; and (d) challenges of defining AI-generated outcome metrics.

---

[10]Miro: https://miro.com

## 5 FINDINGS

This section illustrates the current UX Design tools and workflows, and provides a detailed presentation of the findings under the six themes identified in the data analysis (Sec. 4.3 and Fig. 1). Fig. 3 displays the activities in the UX Design workflow that UX Designers in our study stated can and cannot be completed by GenAI.

### 5.1 Current UX Design Tools and Workflow

Understanding the limitations and challenges of current UX design workflows is crucial for creating a baseline to assess the impact and utility of GenAI tools. This knowledge not only informs how GenAI can address existing issues and enhance workflows, but also aids in contextualizing these tools within different operational environments, from larger enterprises to smaller companies. This is essential for evaluating the integration and potential modifications GenAI tools might bring to these diverse workflows, where one size may not fit all. Moreover, this understanding plays a key role in setting realistic expectations for the adoption of GenAI tools, providing insights into the aspects likely to be embraced or resisted by UX designers.

We synthesized eight categories of frequently used tools that support UX Designers in their daily work (Fig. 2), as well as eight typical steps within UX Design practice – from gathering requirements from business stakeholders and users, to ideation, prototyping, user testing, iteration, and enabling the realization of solutions with the development team. Table 3 illustrates the main activities and tools used in each step. We observe that Figma was mentioned by all participants as a design tool to support high-fidelity prototyping and handoff to development teams. Miro was also frequently mentioned for brainstorming and synthesizing user research data.

In terms of workflow, we observed that larger enterprises (with over 10,000 employees) often have clear divisions of roles between UX Designers, Product Managers, and UX Researchers. In contrast, smaller companies, such as startups with fewer than 50 employees, often lack such clear divisions. UX Designers in smaller companies are responsible for conducting user research and may spend more time convincing stakeholders to prioritize UX. Furthermore, workflows

| Steps | Main Activities | Tools |
|---|---|---|
| 1 Business Requirements | • Analyze stakeholders' requirements<br>• Conduct a stakeholder kickoff meeting to align on expectations, business objectives, and technical challenges. | 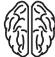 Brainstorm |
| 2 User Requirements | • Define research goals and identify the necessary features to achieve those goals.<br>• Synthesize insights from user research to create personas and empathy maps.<br>• Dive into user's motivation, pain points, and challenges to inform design decisions. | 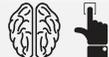 Brainstorm & User Research |
| 3 Ideation | • Convert research insights into design ideas through sketches.<br>• Conduct internal reviews to refine ideas with the design team.<br>• Organize workshops with stakeholders to consolidate design solutions. | 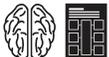 Brainstorm, Documentation & Presentation |
| 4 Prototyping | • Make interactive prototypes that visually and functionally represent the proposed design solutions.<br>• Ensure prototypes capture user interactions, workflows, and navigation paths. | 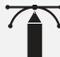 Design |
| 5 User testing | • Conduct user testing to evaluate the prototypes.<br>• Apply online (or offline) moderated (or unmoderated) user testing to guide participants through predefined tasks. | 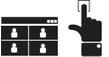 Communication & User Research |
| 6 Iteration | • Analyze user testing results, identifying areas for improvement.<br>• Iterate on the design based on the gained insights | 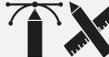 Design & Non Digital |
| 7 Solution | • Collaborate with stakeholders to ensure the design addresses user needs and aligns with the business goals. | 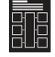 Documentation & Presentation |
| 8 Design Handoff | • Provide the development team with design assets and documentation and guide them to ensure the design's consistency and accuracy across different screen sizes.<br>• Address technical constraints and make UI adjustments to optimize the user interface for the final product. | 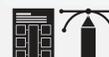 Documentation, Presentation & Design |

Table 3. The typical UX Design workflow in 2023

in smaller companies tend to be shorter than those in larger enterprises due to fewer people involved and lower communication costs. However, smaller companies may also omit or simplify some steps due to resource limitations, such as a limited budget for extensive user evaluations. P6, who has experience working in both small companies and large enterprises, explained, *"In large companies, our [UX Designers'] roles are well-defined. Although we also face a complex business landscape and stakeholders with different opinions, we have collaborators like UX Researchers and Product*

*Managers to handle those aspects, and we can focus more on design. In small companies, we [UX Designers] must take on the roles of researchers, Product Managers, and designers."*

### 5.2 Theme 1: Limitations of Current UX Design Tools

*5.2.1* **Limitation 1: Frictional design handoff to development teams.** Participants pointed out that handing over design specifications, assets, and documentation to the development team is a rather frictional process. The current design tools do not support a seamless transition from design to development in a format accessible and understandable by the developers: "*[We need] more gateways between designers and the developers, improving prototyping workflows by enabling Figma prototypes to feed directly into the development coding workflow, to reduce the frictions in communication between design and development teams.*"[P14]

*5.2.2* **Limitation 2: Fragmented Tools.** Participants emphasized that the current UX Design tools are scattered. Each tool has specific strengths or functionalities, but transitioning seamlessly between them is challenging: "*We use Miro for brainstorming and creating low-fidelity prototypes, and Figma for high-fidelity ones. However, transferring wireframes from Miro to Figma isn't straightforward.*"[P10] Some participants specifically noted that the current design tools lack support for user research: "*Something valuable would be the ability to conduct usability testing and generate data within the design tool itself.*"[P17].

*5.2.3* **Limitation 3: Lack of Advanced Design Support.** Participants expressed dissatisfaction with the current design tools due to their lack of advanced support for complex animations, cross-platform designs, graphic design, and more intelligent design systems: "*There are micro animation-type things that you don't get from Figma, like the timelines that you would use in After Effects or Principle to fine-tune your animations.*"[P17] "*Design systems lack intelligence. I'd appreciate a design system that recommends suitable design components and automatically adjusts these components to fit various screens.*"[P5].

### 5.3 Theme 2: Challenges in Current UX Design

*5.3.1* **Challenge 1: Managing stakeholder expectations and gaining buy-ins for UX..** The most frequently mentioned challenge is that UX Designers often spend over half of their time communicating and collaborating with a variety of stakeholders. This includes working with external vendors who take time to adapt to the work culture and style (P13), spending excessive time keeping stakeholders in sync, coordinating across time zones, and convincing them of the necessity of UX and gaining buy-in from management-level stakeholders (P1, P8, P12). Additionally, they must balance the expectations of stakeholders and clients with user needs to come up with an ideal solution (P18). P12 mentioned, *"I have to convince others that UX is necessary and do a lot of work with developers to ensure the design is appropriately implemented. Some organizations take UX as lightly as a window dressing task and give more buy-in to the backend. So, we need to constantly educate them about the importance of scaling up UX."*. Such management processes, stakeholder communication, and expectation alignment are aspects that heavily rely on crucial human-to-human communication, and an area which GenAI technologies do not currently support.

*5.3.2* **Challenge 2: Understanding end-users.** Participants mentioned that conducting research to understand user needs is challenging in terms of finding out the right pain points of users and building the right products: "*I think the challenge always lies in trying to understand who I'm designing for, what their needs, goals, and problems are, and whether the product or service that I'm working on fits into the broader landscape.*"[P17]. Such a challenge lies at the heart of UX Design and research [12]. While some approaches aim to simulate synthetic users and conversations (e.g., using GPT-3 [74]), it is unlikely this can currently substitute for a deep understanding of end-users by designers.

*5.3.3 **Challenge 3: Designing solutions that satisfy both business value and user needs.*** Participants pointed out that coming up with good design solutions that balance clients' needs, users' needs, business value, and visual novelty is challenging: "*The UX field is quite mature, with existing design references and design systems aiding the work. The challenge is to come up with new ideas constantly and find a balance when considering both established design references, internal stakeholders' expectations, user needs, and incorporating them into the new design.*"*[P18]* Satisfying business needs can carry implications for originality, copyright, and ownership, all of which impact adoption (cf., Sec 5.6.1).

*5.3.4 **Challenge 4: Lacking domain knowledge and resources.*** Participants mentioned that lacking domain expertise and resources to access knowledge poses a significant challenge: "*One of the challenges for UX Designers is acquiring extensive domain knowledge, such as creating designs customized for specialized fields like medical, automotive, fintech, and so on.*"*[P1]*; "*In a startup, we often face resource constraints that limit our access to experts, conducting user research or usability testing on a larger scale, which is crucial to ensure the feasibility and desirability of the products.*"*[P6]*. We believe this may be a promising avenue by which GenAI tools, specifically text-based large language models such as ChatGPT, can empower designers. Despite current model hallucinations (cf., [46]), designers can have immediate, vast, and easy access to domain-specific information and resources.

## 5.4 Theme 3: UX Designers' Attitudes toward GenAI

All participants agreed that GenAI tools can be powerful and convenient in supporting general tasks such as early-stage design ideation, basic UX writing, and basic coding. These tools provide a starting point for human designers to build upon, helping them overcome the fear of beginning from scratch. While many participants were optimistic about the potential of GenAI tools, envisioning these tools as a means to enhance design efficiency, some others expressed mixed feelings towards GenAI. On one hand, they acknowledged that AI could assist with repetitive tasks, but they underscored that UX Design is inherently user-centered. The reliability of AI-generated designs is questionable, as they may not always align with user needs and goals. Relying too heavily on AI to design products or interfaces without human input was seen as risky. All participants emphasized that the current quality of AI outputs, such as text accuracy, image resolution, and design details and functionalities, still requires significant human input and review, which may not necessarily save time.

All participants believed that human designers possess unique abilities, as they exhibit empathy towards users and derive enjoyment from the design process, even when starting ideation from scratch: "*GenAI tools can help nowadays in generating basic ideas to help us populate the blank canvas, thereby aiding in overcoming the fear of the 'empty canvas'. It's similar to when we need to write reports; you can start with a basic structure generated by ChatGPT and then build upon it. However, I fear that in the long run, I might become more of an editor, gradually losing my design superpower to fill in the blank canvas and be empathetic towards users.*"*[P8]*.

*5.4.1 **Perceived shortcomings of GenAI tools shown in the demo***. There were two shortcomings perceived to be key amongst participants:

**(1) Practicality and efficiency**. Participants highlighted that the GenAI tools showcased in the demos are quite generic, and the design process still requires a substantial amount of manual work. This includes tasks such as redrawing icons and buttons and inputting accurate text-based prompts. "*Designers are trained to think visually. Our design process *involves trial and error on sketchbooks. Designs emerge from numerous sketch trials, not from text-based prompts. [In the demo], you're required to switch between various GenAI tools, Photoshop plugins, and perform a substantial amount of manual work, including redrawing and regenerating high-resolution images. I question whether these GenAI tools truly enhance the design workflow."[P12]*

**(2) Accuracy and reliability.** Participants questioned the overall accuracy and reliability of outcomes generated by AI tools. They hold the belief that human inputs are necessary to validate the AI-generated outcomes. P16 shared: "*For visual design, GenAI tools are still quite limited in capturing nuances and intricate details. Regarding ideation, AI tools like Miro AI may fall short in adequately representing the insights of stakeholders who possess extensive experience, say, 20 years in a specialized field like healthcare."[P16]*. P19 described her negative experience using GenAI tools: "*For Midjourney, I often found myself spending hours adjusting my prompts, yet still unable to achieve the envisioned results. As for Miro, the cards are only clustered based on keywords, requiring manual effort to create coherent clusters."[P19]*. Furthermore, P3 expressed doubts about the reliability of AI's code generation capability: "*I've actually attempted to have ChatGPT generate code for certain features or functionalities, but it didn't quite produce the results I was looking for."[P3]*.

## 5.5 Theme 4: Human-AI Collaboration

All participants described GenAI tools as helpers or assistants and recommended that "augmentation" and "enhancement" are good words to describe the assistance provided by GenAI in UX Design.

### 5.5.1 *Perceptions of threats and collaboration with GenAI*.
Participants expressed varying levels of concern regarding the potential threats of GenAI. Some felt that if GenAI remained a generic tool under their control, it wouldn't be a source of concern, while others worried about its ability to simulate designers' reasoning. Recommendations included treating GenAI as a team member, with some participants emphasizing the importance of humans retaining ultimate decision-making power in human-AI collaboration. P18 commented: "*If the technology does eventually reach a point where it can accurately simulate designers' rationales, then I might start worrying about it."[P18]*. P19 recommended treating GenAI as a team member: "*I would envision AI as part of the team, contributing to the overall process, but not solely responsible for the final outcome."[P19]*. P8 stressed that humans should have the ultimate decision-making power in human-AI collaboration: "*It should depend on people to choose which part I want AI assistance, for the purpose of expediting tasks. I don't like the word 'replace'. I very much enjoy the UI and UX Design...I want AI to be my apprentices. Humans do the creative and complex parts, and AI apprentices fill in the colors or draw lines."[P8]*.

### 5.5.2 *General and repetitive work can be replaced by GenAI*.
All participants shared the belief that GenAI has the potential to replace repetitive tasks, such as generating "lorem ipsum" texts for design mockups, condensing design briefs and technical documentation, crafting repetitive design components, creating responsive designs for various screens, automatically documenting visual design elements into design systems, generating low-fidelity prototypes, and generating design variations for inspiration (see Fig. 3: "GenAI Can"). As P8 said, "*I envision AI functioning as a productivity tool that quickly aids me in concrete tasks. For instance, AI could summarize PRDs [Product Requirement Documents] or explain technical backgrounds in plain language. Currently, these aspects require substantial effort from PMs [Product Managers] and engineers to convey to us designers."[P8]*. P12 expressed their wish for AI-supported responsive design, "*I'm hopeful that GenAI can assist with tasks like responsive design for different screens and suggesting user flows across pages, not just on a single page."[P12]* Many other participants expressed their desire for GenAI support in various tasks, including providing design inspiration (P2, P4, and P8), as well as facilitating efficient UX writing (P18 and P19).

P19 shared: "*In our startup, we don't have a dedicated UX writing role. Our designers often use ChatGPT to assess the appropriateness of the UX content in our design.*"[P19].

### 5.5.3 UX Design work cannot be replaced by GenAI.
Many participants made it clear that much UX Design work simply cannot be replaced by AI, including the work that requires communicating between human stakeholders, being empathetic towards user needs, and designing user-centered functionalities. They stressed the importance of humans being the original creator and validator of the AI outcomes (see Fig. 3: "GenAI Cannot").

**(1) User inputs and human-human communication or collaboration.** Participants confidently expressed that they do not currently perceive GenAI as a competitor, since human designers serve as ambassadors for user needs and continue to be the ultimate decision-makers and arbiters of AI alignment. They emphasized that GenAI cannot replace work that requires user inputs, and human-human collaborations: "*GenAI might be capable of taking over basic, repetitive, and straightforward tasks, but it cannot replace service design and collaborative efforts that demand more systematic and abstract thinking.*"[P6]. P14 additionally pointed out, "*UX Design is not just about aesthetics, but also usability and the accessibility of elements. Real human users are essential for improving designs through actual user research and testing.*"[P14]. P16 believed that human communication is not replaceable by AI, stating "*I value insights from human colleagues, their experiences, ideas, and the knowledge they bring to the table. These do not have to be flawless, but they aid us in charting directions together. I do not believe AI can perform this aspect of work.*"[P16].

**(2) Human creativity and decision making.** Some participants pointed out that human creativity and originality are strong human qualities that are difficult to be replaced by AI. AI outcomes heavily rely on past data, potentially leading to recurring results due to the limited data pool. Human designers remain the driving force capable of deeply comprehending design contexts, empathizing with users, and innovating based on the latest knowledge: "*The creative work that humans enjoy doing should not be replaced by AI. AI is a trained model that is based on the past, while human creativity and originality are more future-oriented.*"[P8]; "*Creativity can be supported by AI, as it can remix and regenerate based on human-inputted creative data, but humans possess the originality.*"[P4].

Several participants stressed that final decision-making cannot be delegated to GenAI. Human verification is deemed necessary for each stage of design and development projects: "*AI must be intentionally used, monitored, and validated by humans. They are essentially Large Language Models, which are computers trained to converse like humans. The quality of the content and data provided by AI is not necessarily supported by actual research data verified by human researchers.*"[P8].

## 5.6 Theme 5: Ethical Concerns and Ownership

Copyright, privacy, data biases and transparency, skill degradation and unemployment, social disparity, and equal access are frequently mentioned ethical concerns among all participants.

### 5.6.1 Originality, copyright, and ownership.
Participants emphasized the significance of crediting the original creators of artworks or art styles utilized in UX work. P3 and P4 suggested that blockchain technology could be useful in ensuring direct ownership attribution to the original creators, regardless of the extent to which the original work has been modified. P4 stated, "*Ownership must remain traceable regardless of how extensively the AI outputs have been modified, derived, or reproduced from the originals. Perhaps blockchain can provide a solution here.*" P14 expressed her concerns over design plagiarism, "*From the confidentiality point of view, what data can be placed into GenAI tools, how we can make sure that we are using legal sources and we are not violating GDPR regulations. There are already cases that designers copy others' work to build their own portfolios. GenAI can make plagiarism in design much worse.*"[P14].

Expanding on the subject of ownership, some participants believed that humans must be the owners, with GenAI tools serving as supportive aids to humans in achieving the final results. P11 commented, "*AI is not human; ownership belongs to humans. AI is a tool, much like Figma. I don't believe ownership should be assigned to technology or tools. If you are the one refining, crafting the prompt, contributing creativity and thoughts to AI for generating the work, and overseeing the entire process, ownership should be attributed to humans.*"[P11].

However, several other participants recommended attributing ownership to the part of work supported by AI or labeling it as AI-contributed and ensuring transparency about the tools used. P18 provided a fitting analogy, "*The ownership attribution for using AI in future work should be similar to how researchers present their research findings, mentioning the sources and systems used as references.*"[P18] This was echoed by P19, "*Ownership should be seen as a collective effort with input from various team members, including GenAI.*"[P19].

5.6.2 **Skill degradation and unemployment**. Several participants highlighted their concern that human UX Design work might be less appreciated, potentially leading to complacency among human designers and a decline in their skills. They worry that AI taking over tasks previously considered essential for designers could contribute to skill degradation. These concerns, mentioned primarily by senior designers (refer to Table 1), are particularly relevant for junior designers, who could miss out on opportunities to further develop their skills. As such, human designers might transition into more managerial or generalist roles. P16 (a senior designer) envisioned this, stating, "*Designers may fall into the trap of becoming complacent. Junior designers might have fewer chances to learn the design process but, instead, they may focus on mastering AI systems. Consequently, designers could evolve into generalists.*"[P16]

Participants specifically highlighted that unemployment and skill degradation could evolve into significant societal issues if not addressed seriously from the outset. P4 (junior designer) and P5 (senior designer) expressed concerns that human designers' skills might become narrow as we increasingly specialize in specific tasks if AI potentially takes over many parts of the work in the design workflow, potentially limiting our creativity as designers: "*AI can put us into a bubble. We feel that we are doing higher-level work, and the career barrier of becoming a designer is getting higher. But somehow we are forced to be more specialized in a limited range of tasks. The loss of oversight over the design and development workflow may jeopardize our creativity.*"[P5].

P12, also a senior designer, thoughtfully reflected on the impact of GenAI on education and careers of junior designers, "*Our expertise matures through year-over-year practice, encompassing not only design skills but also design thinking and genuine user empathy. If companies decide to employ prompters or AI operators in the future to replace junior designers' roles, how can these prompters or AI operators receive the proper training equivalent to what we receive in design schools or through design practice? More importantly, how can junior designers acquire new skills or senior designers improve their expertise if AI takes over the tasks they need to practice and uphold their design skills?*"[P12].

5.6.3 **Privacy, transparency, and equal access to GenAI**. Participants pointed out that privacy is a significant concern within the realm of GenAI. As these AI models ingest large amounts of data to learn patterns and generate content, there is potential for the unintentional exposure of sensitive information, such as personal or confidential details that users did not intend to share: "*If I want to use ChatGPT to extract information from technical documents, I don't know how I should prevent sensitive information from being leaked. For my personal life as well, it is not clear how much of my personal data will be retained to further train the AI models.*"[P19]

Several participants stressed the importance of AI being trained on unbiased data and ensuring equal access to GenAI tools. This is due to concerns about social disparity and the potential for country policies to limit people's ability to learn and access AI tools, as well as the possibility of biased data leading to questionable outputs. "*Big companies and *developed countries may gain control over the data fed into AI. It's also easier for them to access GenAI tools compared to some developing countries."*[P6]. P2 added, *"It is important to ensure that individuals who are relatively less educated and have limited knowledge of GenAI can still use these tools to enhance their work and daily lives."*[P2].

### 5.7 Theme 6: The Future of GenAI in UX Design

Theme 6 provided insights into participants' visions and desires related to the integration of GenAI in the field of design.

*5.7.1* **Envisioning collaborative design with GenAI.** Some participants envisioned a triangulated system that involves collaboration and input from users, designers, and GenAI. This approach aims to minimize biases and ensure that crucial nuances, essential for design innovations, are not overlooked. As P11 mentioned, *"Human designers may overlook certain nuances that could hold importance for users. I wish for GenAI to enhance this aspect by providing its input grounded in vast past data and knowledge, potentially elevating user experience design to the next level."*[P11].

*5.7.2* **Promoting AI literacy and access for designers.** Several participants expressed a hope for increased AI literacy through education, benefiting both experienced and junior designers. P13 (a junior designer) noted, *"Students may feel apprehensive about this significant AI revolution. Education is essential to prepare both junior and senior designers for the impact of AI and how to leverage it to enhance our human design skills."*[P13]. Additionally, a few participants envision future AI tools to be more accessible for designers. In other words, they desire AI tools that offer greater visual and intuitive usability, moving beyond sole reliance on text-based inputs. *"The recent Apple Vision Pro headset is inspiring. Its visual and intuitive design appears promising. Thus, AI should also evolve to become more visual, intuitive, and universally accessible to all users."*[P12].

*5.7.3* **AI's desired role in enhancing design processes.** Many participants expressed their desire for GenAI to mature further, supporting them in various design tasks. They hoped that GenAI could assist in designing graphics, writing prompts, generating smart components, providing successful design examples, and automating the handoff of designs to development teams. P4 noted, *"I am anticipating easier ways to write prompts or automatically generate components that adapt to different screens, recommend smart animation transitions, and even generate 3D content for immersive environments in the future."*[P4]. Many other participants also voiced their desires for specific GenAI support in various tasks in the future. For instance, they hoped AI could summarize state-of-the-art knowledge for research and strategy (P10, P18, and P19), serve as a synthetic user to expedite sprints (P1 and P18), offer professional guidance in accessibility design (P12), help designers quickly master and switch between different design tools (P18), aid in labeling and clustering qualitative raw data from user studies (P5, P9, P10, P12, and P18), and seamlessly translate designs into development (P5, P18, and P19).

P18 specifically highlighted the constantly changing landscape of design tools, expressing a desire for *"GenAI to reduce the time designers spend keeping up with rapidly evolving technology. Design tools change every few years. If AI could help designers transition seamlessly between old and new tools or consolidate everything into a single tool, designers would have more time to understand user needs and design."*[P18]. P5 expressed a wish for *"GenAI tools to help me assess the feasibility of designs on the development side during the early design stage and assist in documenting designs in Figma, facilitating smooth handoffs to the development team in a format accessible and understandable to developers. Currently, I still have to manually document designs, take numerous screenshots, and verbally explain them to the development team."*[P5].

*5.7.4* **The challenge of defining GenAI outcome metrics**. According to several participants, another envisioned future for GenAI involves the challenge of defining metrics to measure the quality of AI-generated output. As P17 commented, "*If many businesses use the same GenAI tools, it may result in similar designs and lacking diversity and innovation. Also, AI can lead to generating numerous outputs and leave the challenge to us, humans, to choose among them.*"[P17]. "*It is already a challenge to select and fine-tune AI outcomes. We need guidelines on how to measure or validate AI's output.*"[P18]

## 6 DISCUSSION

Our key findings indicate that experienced UX Designers acknowledge GenAI's potential as an assistive tool for repetitive tasks and productivity enhancement, yet they remain cautious about adopting GenAI tools themselves. This is due to their limitations in understanding user-centered design and their potential impact on their creative skills, as well as fundamental limits to enjoyment and agency in human-AI collaboration. While GenAI is seen as a game-changer in addressing challenges within current UXD practice, such as friction in design handoffs and limitations in UXD resources, designers expressed grave concerns about skill degradation, particularly for junior designers. They advocate for GenAI tools to evolve into intuitive, visually-oriented, and collaboratively intelligent systems. This evolution aims to balance human creativity and AI efficiency, emphasizing responsible AI use and human-AI collaboration to preserve human control in creativity and decision-making. We delve into these aspects below, where we discuss implications for human-GenAI collaboration, specifically copyright and ownership, human creativity and agency, and AI literacy and access.

### 6.1 Enjoyment and Agency in Human-GenAI Collaboration

Enjoyment and meaning are two recurring key words in our conversations with UX Designers. As UX Designers not only aim to excel in their work but also seek to derive pleasure from it. Recent work has found that the user experience of early users of ChatGPT was impacted by hedonic attributes, not just pragmatic ones [88]. Enjoyment in work can be viewed through the lens of motivation theory, including both extrinsic and intrinsic motivations [48]. Extrinsic motivations revolve around whether the technology efficiently serves as a means to accomplish tasks, treating technology such as GenAI as a tool. Intrinsic motivations center on whether UX Designers find their work inherently rewarding and the engagement with technology enjoyable, considering technology such as GenAI as a toy [8]. We argue that, in UX Design, where creativity and problem-solving play central roles, this intrinsic motivation becomes paramount. Designers are not just driven by the end results; they also seek fulfillment and enjoyment throughout the design process. In other words, the future of AI-human collaboration should empower individuals to focus on tasks that truly matter to them, allowing AI to handle more mundane aspects. This approach not only enhances productivity but also fosters job satisfaction [59].

Closely related to enjoyment, we found that agency in human-AI collaboration is also a recurring theme. Will AI take on more agency while humans play a supervisory role? Or will humans retain primary control and have the ability to override AI's decisions when necessary? Alternatively, will AI and humans collaborate in a mixed-agency approach, where both contribute to decision-making and execution [91]? UX designers believe that we should differentiate between well-defined, routine tasks and complex, creative problem-solving tasks. Humans bring contextual understanding, ethical judgment, and accountability, and should lead complex work that requires creativity, empathy, ethics, and complex decision-making. Tasks led by AI and decisions made by AI should be transparent (cf., deceptive AI ecosystems [105]), explainable [21], and easily adjustable by human collaborators. The ideal scenario involves humans learning from AI insights, while AI improves based on human feedback (see e.g., [30] for a review on challenges for Reinforcement Learning from Human Feedback). Overall, we found that UX designers' beliefs align with those of Fanni et al. [26], who advocate for "active human agency" and argue that AI policies should include provisions that enable users to promptly challenge or rectify AI-driven decisions (cf., contestable AI [66]). This approach aims to empower individual autonomy and bolster fundamental human rights in the digital age. We find that the foregoing dimensions of enjoyment and agency to not only be unique human factors, but we suspect will increasingly play a role in the extent that UX Designers retain usage of such tools.

### 6.2 GenAI Interpolating, Humans Extrapolating, and Creativity Exhaustion

In today's efficiency-driven agile working culture, UX Design is essential for getting the design right and the right design [12]. It brings together a team to go through each sprint, progressing from idea to prototype to user research within short time frames [18]. GenAI's remarkable speed in generating content and solutions have the potential to facilitate the fast-moving sprints. However, we must be aware that while GenAI can produce vast quantities of outputs in an instant, humans are the innovators behind these creations. Essentially, AI is interpolating, estimating values or generating outcomes within the range of existing data. Thoring et al. [93] approach this from the lens of evolutionary creativity theory [87], where Variation and Selection are key components. While GenAI interpolates (Variation), it currently lacks the capability of extrapolating out of the latent space and continuing into unknown territory or merging into another latent space without supervisory human control (Selection) [93]. Indeed, AI might excel at "peripheral work" [83] or repetitive tasks within known data patterns. However, since AI is trained with known data, relying heavily on AI can lead to homogenized solutions (e.g., the AI-generated user interfaces all look similar) [6] or amplify biases in the data (e.g., the resulting design favoring certain demographic groups [4, 19, 25]. In contrast, consider children, who receive around four or five orders of magnitude less language data than LLMs, and vastly outperform capabilities of AI [31]. Indeed, we humans excel at extrapolating, making predictions beyond the known data range using not only our creativity, intuition, and domain knowledge, but also benefit from multimodal grounding and the social and interactive nature of received sensory input [31]. In essence, humans excel at the "core work" that they identify with, which contributes to not only their success, but also their happiness [83].

Another potential setback revolves around the danger of human creativity exhaustion in keeping up with the pace of GenAI output. GenAI's relentless speed can exert immense pressure on human creators. While AI is not directly subject to physical factors like fatigue [3], humans are expected to innovate continuously and rapidly on a large scale. On one hand, AI's speed sets a challenging standard for human creators; on the other hand, it also lowers the training requirements to become an artist or a creative, as many people can create artistic paintings without picking up a brush. Moreover, as AI learns from human creations and refines its abilities, it raises the bar for what is considered innovative. Human creators in general, and UX Designer in particular, may feel compelled to push their limits to compete with AI, which can be mentally and emotionally draining. To sustain human creativity, instead of competing with AI's speed, human creators seem to be better off focusing on tasks that require critical thinking and emotional depth, where the human touch is irreplaceable. To that end, support systems and education should be introduced to alleviate the pressure on humans and help mitigate the risk of creativity exhaustion, a point to which we discuss next.

### 6.3 AI Literacy and Participatory AI in UX Design

In UX Design, there is a growing presence of tools with AI features or supporting AI plugins. AI is gradually being integrated into various stages of the UX workflow [65]. AI literacy begins with a fundamental understanding of AI concepts, including how machine learning works, the types of AI systems, their capabilities, limitations, and implications. This understanding is crucial for UX designers to harness AI's power, ensure equal access to AI education and tools, and actively participate in shaping and refining AI technologies, especially those used in their design processes. By adopting a participatory AI approach, UX Designers are part and parcel of the conversation around the future of GenAI development – this enables UX designers to strengthen their feedback and expertise, ensuring that AI tools align effectively with human intentions and UX design practices [57, 63].

Another core element of AI literacy is that UX Designers should be able to think critically and question AI-generated outputs. There is a need for a comprehensive framework or guidelines for evaluating AI outcomes. Such a framework could encompass the assessment of AI's credibility, performance, accuracy, and impact on user experiences [89], as well as provide a structural means for contestability [66]. The foregoing echoes recent efforts [50] to ensure committees consisting of AI experts, ethics experts, and importantly creative professionals, are brought together to oversee GenAI's progress. These groups would be best equipped to provide standards and best practices for evaluating GenAI output and mitigating harm on creative professionals (cf., [50]).

### 6.4 Copyright and Ownership for GenAI Output in and beyond UX Design

In our study, UX Designers emphasized the significance of crediting the original creators of artworks or styles utilized in AI work. They mention the use of blockchain technology (cf., [24]), where of special interest are computational provenance mechanisms [94] that track and document the origins, processes, and transformations of AI-generated content or artifacts. As such, this involves recording information about how a piece of content, such as text, images, or other media, was generated, which AI models and algorithms were used, and any data sources that contributed to the creation of the content. In so doing, this would facilitate the attribution of ownership should they draw on model output throughout their UX practice. We found that many UX Designers in our study believed that humans should be the owners of the design outcomes. However, this was so only if GenAI tools serve as supportive aids, and human designers remain in the driver seat by leading the entire process, adding detailed touches and synthesizing user inputs. However, they recommended attributing ownership to the part of the work supported by AI or labeling it as AI-contributed while ensuring transparency about the tools used. Interestingly, in a recent study, Draxler et al. [20] found that people often do not disclose AI as an author when using personalized AI-generated texts in writing, although they attribute ownership to the AI. Comparing human and AI ghostwriters, they found that attributing authorship to oneself is more prevalent when writing with AI support. Given disagreement in attributing ownership, it is necessary to develop an authorship attribution framework that takes the user and their relationship to the generative model into account.

The discourse around copyright, privacy, ownership for UX design practice is inherently connected to the wider discourse on GenAI as a whole. Whether dealing with text, images, or video, the foregoing raise the question of whether the GenAI output being generated is ethical and legally compliant to begin with. Indeed, recent work by Jiang et al. [50] in assessing the impact of visual GenAI on artists paints a bleak picture regarding the outspoken harms on artists, which can well extend to other creative works (including the work of UX designers). For example, "fair use" within US copyright law may not always end up protecting the artist, whether due to case-by-case determination, or the high cost in pursing legal battles. Whether the way forward constitutes better watermarking approaches to ensure copyright and traceable accountability [94, 107], or the further development of regulation and policy at an (inter-)national level (e.g., European AI Act[11] [97]), it is clear GenAI development needs to be acted upon responsibly - for and beyond

---

[11]https://eur-lex.europa.eu/legal-content/EN/TXT/?uri=CELEX:52021PC0206

UX Design practice. As such, we believe drawing on the key human-centered principles (e.g., aligned with human well-being, respecting privacy, and having appropriate governance and oversight) of responsible AI development [34] are of immediate necessity to protect designers moving forward.

### 6.5 The Future of GenAI-infused UX Design: Fears and Opportunities

The rapid advancements in GenAI and UX Design underscore the necessity of ongoing monitoring and research in both fields. As GenAI tools continue to evolve, it is essential to ensure that they align with human values, ethics, and creative processes, despite any uncertainty that may arise from currently unresolved empirical and conceptual issues in human psychology [11]. To that end, by answering our research questions of current **(RQ1)** and future **(RQ2)** perceptions of GenAI, this study serves as a foundation for exploring the opportunities and challenges presented by GenAI in UX Design, where we emphasize the importance of responsible and informed collaboration between humans and AI in this domain, which leads to the question: are such fears of GenAI for UX Design warranted?

We observed (Sec. 5.5) that experienced UX Designers' were defensive about the very idea that "AI can potentially replace human designers". Experienced designers, on the one hand, have confidence in their superpowers in terms of design skills, the collision of wisdom and enlightenment in human-human communication, and possessing high empathy towards users. However, they emphasized (Sec. 5.6) the real risks that GenAI could pose on the junior UX Design workforce, posing the dilemma: if junior human designers don't do the mundane work that can essentially be replaced by GenAI, then how will they reach the seniority levels wherein GenAI cannot replace? On the other hand, even experienced designers raised fears about AI in general, and GenAI in particular, and how this may in the future render their expertise partially obsolete. Specifically, some concerns revolved around emergent AI capabilities, where as AI becomes more sophisticated (e.g., more parameters with more diverse and larger training data), there is the possibility of it advancing beyond the intended scope or control [6, 80]. On the flip side, some of our participants may not have been aware of the rapid pace of developments – for example, GenAI models can be fine-tuned on unique stakeholder datasets [104], which would elevate ideation, brainstorming, and documentation within the UX design process. Nevertheless, the risks persist should even fine-tuned models lead to situations where GenAI systems generate designs or make decisions that are not aligned with refined user experience principles, stakeholder satisfaction, or worse, violate human values, ethics, or cultural norms [100, 108], even when designers may take the driver seat of selecting and refining the output with additional prompts (cf., [15]) on custom-trained models. These concerns align with user perceptions of machine consciousness (of GPT-3) [84], which lead to several tensions, of which most relevant to our use case are interaction and experience, control and independence, and rigidity and spontaneity. Furthermore, as in digital art, people who exhibit strong anthropocentric creativity beliefs may be biased to brush off GenAI design output as overall less creative and awe-inspiring should they know it was made with GenAI [68].

However, unpredictability, polarity, and such tensions often signify new opportunities as emerging AI capabilities can lead to innovative problem-solving approaches and entirely new design concepts that while inspire human creators, can radically disrupt the practice (reminiscent of paradigmatic shifts in the natural sciences [54]). However, while we already see a radical disruption to several elements of UX Design practice (cf., Fig. 3), much remains to be desired (and cautious about) as GenAI capabilities increase. Essentially, we contend that ensuring harmonious alignment between human-AI collaboration for UX Design practice, methods and strategies are required to ensure that GenAI systems comprehend and respect human designers' values and interpret and respond to their concerns effectively. For this, we call upon further research in and beyond responsible AI to continue placing the human at the center of technology development, and importantly, on reconciliation efforts should ethical values differ amongst groups [49]. Indeed, as Rakova et al. [78] argue, to better enable responsible AI work, organizations need to update their prevalent practices, which requires addressing prevalent work practices, emerging work practices, and mapping their aspirational future. To this end, a path forward is to continue drawing on participatory AI and value-sensitive design [5, 21, 33, 47] to create AI for social good [95], and ultimately foster human and creative safeguarding mechanisms, allowing UX Design practice and its practitioners to adapt and flourish gracefully.

### 6.6 Study Limitations and Future Work

In our investigation, we opted for a breadth-first approach to better understand the overall impact that GenAI may have on UX Design, which naturally comes at the cost of depth – we did not delve deeper into any particular UX products, tools, or workflows. As such, prioritizing breadth over depth was strategic, enabling the capture of a wide range of perspectives and considerations within the UX field. This serves as an important stepping stone for future works to delve into specific tools and use cases to provide a more in-depth analysis. Second, our interviews were limited by the small sample size (N=4) of startup participants. Startups often have unique team structures and resource constraints, leading to different approaches to addressing UX Design challenges compared to established companies. This invites future work to study the organizational practices of startup ecosystems, and how GenAI is embedded within such structures. Furthermore, although the majority of our participants had experience using GenAI tools such as chatGPT (GPT-3.5) and Midjourney, the study invited participants to watch demonstrations of GenAI tools without actual hands-on usage. This limited interaction might have overlooked practical insights that could arise only from direct hands-on experience with such tools.

## 7 CONCLUSION

Our work set out to investigate how UX Designers perceive GenAI tools, and how this is reflected in their current workflows and practices. Further, we identified the opportunities and risks do UX Designers envision for the future of human-AI collaboration in UX Design practice. Through in-depth interviews with 20 UX Designers who have diverse years of experience, we found that experienced designers are confident in their originality, creativity, and empathic skills, and find GenAI's role overall as assistive. They emphasized factors of "enjoyment" and "agency" as uniquely humans, where humans will always remain in the driver seat over AI output. However, there were serious concerns over setbacks for junior designers, who may be impacted by skill degradation, job replacement, and creativity exhaustion. We draw several implications for responsible human-AI collaboration in UX design, specifically related to copyright and ownership, human creativity, and AI literacy and access. We call on continued efforts to develop responsible AI for ensuring human and creative safeguarding mechanisms, allowing UX Design professionals to immediately yet gracefully adapt to the dawn of GenAI, and its disruptive impact on UX Design practice.

## ACKNOWLEDGMENTS

We thank all our participants for taking part in our study, and offering us their valuable perspectives on GenAI.

User Experience Design Professionals' Perceptions of Generative AI				CHI '24, May 11–16, 2024, Honolulu, HI, USA